\documentclass[twocolumn,aps,prl,showpacs]{revtex4}
\usepackage{graphicx}
\begin{document}

% Use the \preprint command to place your local institutional report
% number in the upper righthand corner of the title page in preprint mode.
% Multiple \preprint commands are allowed.
% Use the 'preprintnumbers' class option to override journal defaults
% to display numbers if necessary
%\preprint{}

%Title of paper
\title{Influence of Carbon Concentration on the Superconductivity in
MgC$_x$Ni$_3$}

\author{L. Shan$^1$, K. Xia$^2$, Z.Y. Liu$^1$, H.H. Wen$^1$, Z.A. Ren$^1$, G.C. Che$^1$, Z.X. Zhao$^1$}

\affiliation{ $^1$National Laboratory for Superconductivity, Institute of Physics,
Chinese Academy of Sciences, P.O. Box 603, Beijing 100080, China}

\affiliation{$^2$International Center for Quantum Structures, Institute of Physics,
Chinese Academy of Sciences, P.O. Box 603, Beijing 100080, China }

\date{\today}

\begin{abstract}
% insert abstract here
The influence of carbon concentration on the superconductivity (
SC ) in MgC$_{x}$Ni$_3$ has been investigated by measuring the low
temperature specific heat combined with first principles
electronic structure calculation. It is found that the specific
heat coefficient $\gamma_n=C_{en}/T$ of the superconducting sample
($x\approx1$) in normal state is twice that of the
non-superconducting one ($x\approx 0.85$). The comparison of
measured $\gamma_n$ and the calculated electronic density of
states ( DOS ) shows that the effective mass renormalization
changes remarkably as the carbon concentration changes. The large
mass renormalization for the superconducting sample and the low
$T_{c}$( 7K ) indicate that more than one kind of boson mediated
electron-electron interactions exist in MgC$_{x}$Ni$_3$.

\end{abstract}

% insert suggested PACS numbers in braces on next line

%\maketitle must follow title, authors, abstract, \pacs, and \keywords
\maketitle

% body of paper here - Use proper section commands
The recently discovered superconductor MgCNi$_{3}$ \cite{HeT2001}
with an anti-perovskite structure has attracted renewed attention,
because it is a new intermetalic compound following the remarkable
discovery of MgB$_2$ \cite{NagamatsuJ2001} and is thought to be
very close to ferromagnetism (FM) due to a large proportion of Ni
in each unit cell \cite{RosnerH2002}. Experiments have shown that
the superconductivity (SC) in MgC$_x$Ni$_{3}$ is very sensitive to
the content of carbon and disappears below about $x=0.88$
\cite{AmosTG2002}. Moreover, Ni-site doping with Cu (electron
doping) and Co (hole doping) have very different
effect\cite{HaywardMA2001}.

Many efforts have been focused on the origin of SC in MgCNi$_{3}$.
Some researchers explored the possibility of conventional
phonon-coupled pairing \cite{ShimJH2001,DugdaleSB2001}. However,
various experiments and theoretical calculations have not given a
consistent coupling constant $\lambda$ in the frame of
electron-phonon coupling (EPC) mechanism. For example,
$\lambda\approx1-1.6$ was estimated from the calculated plasma
frequency ($\omega_p$) and measured resistivity
\cite{SinghDJ2001}, while the measured upper critical field
$H_{c2}$ \cite{LiSY2001,WalteA2002,MaoZQ2002} suggested a much
larger value of $\lambda\approx2.5-3$. Furthermore, two discrepant
values of $\lambda=0.7$ \cite{HeT2001} and 2.2 \cite{WalteA2002}
were determined in the similar specific heat measurements by
taking different approximations. Besides these confusing results,
it has been proposed that EPC is possibly not the only
contribution to the coupling mechanism in MgCNi$_{3}$ system
\cite{SinghDJ2001,WalteA2002}. Opposed to the conventional pairing
mechanism, it was suggested that FM and SC may coexist due to the
ferromagnetic instability caused by the high density of electronic
states (DOS) at the Fermi energy (E$_F$)
\cite{RosnerH2002,SheinIR2001}. NMR experiment confirmed the
presence of substantial ferromagnetic spin fluctuations in
MgCNi$_3$ \cite{SingerPM2001}, while it also exhibited a clear
coherence peak of the nuclear spin-lattice relaxation rate just
below $T_c$, which is typical for an isotropic s-wave
superconductor. More recently, Voelker {\it et al.}
\cite{VoelkerK2002} proposed that MgCNi$_3$ could be a multiband
superconductor with a conventional phonon mechanism
\cite{AgterbergDF1999} similar to MgB$_2$ \cite{ChoiHJ2002}. Thus,
the origin of SC in MgCNi$_3$ is still controversial. One of the
most illuminative methods is to study the influence of
element-substitution on SC.

In this Letter, we compare the specific heat data of a superconducting MgC$_x$Ni$_3$ with
$x\approx 1$ and that of a non-superconducting one with $x\approx0.85$. Using the
tight-binding linear muffin-tin orbital (TB-LMTO) band method and coherent potential
approximation (CPA) \cite{AndersenOK1985, TurekI1997}, we investigate the carbon
concentration dependence of the electronic structure of MgC$_x$Ni$_3$. Combining the
results of the specific heat and band calculations, we propose a self-consistent picture
to understand the reduced SC in non-stoichiometric MgC$_x$Ni$_3$.

\begin{figure}[top]
\includegraphics[scale=1.1]{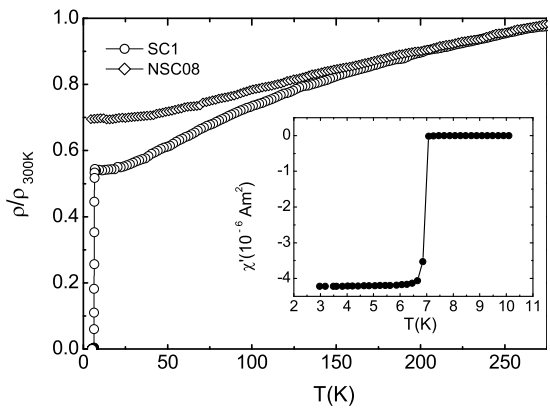}
\caption{\label{fig:fig1}The temperature-dependent resistivity (normalized by the value
at 300K) of the samples SC1 and NSC08. Inset: The enlarged view of the superconducting
transition in AC susceptibility measurement. }
\end{figure}

Poly-crystalline samples of MgC$_x$Ni$_3$ were prepared by powder
metallurgy method. Details of the preparation were published
previously \cite{RenZA2002}. The superconducting transition of the
stoichiometric sample( SC1 ) with $T_c = 7K$ is shown in
Fig.~\ref{fig:fig1}. A non-superconducting sample (NSC08) was
synthesized by continually reducing the carbon component until the
diamagnetism was completely suppressed. X-ray diffraction (XRD)
patterns show nearly single phase in both samples. The lattice
parameters $a$ determined from XRD are 3.812$\AA$ and 3.790$\AA$
for SC1 and NSC08, respectively. By comparing these lattice
parameters with the reported $a\sim x$ relation \cite{AmosTG2002},
we could estimate the carbon content as $x=0.977$ and 0.850 for
SC1 and NSC08, respectively. According to the $T_c\sim x$ relation
presented in Ref \cite{AmosTG2002}, $x=0.977$ corresponds to
$T_c\approx7K$ and $x=0.850$ is very close to the critical
stoichiometry of SC in MgC$_x$Ni$_3$, which is in good agreement
with our experimental results. The temperature dependence of the
normalized resistivity is shown in Fig.~\ref{fig:fig1}. The heat
capacity data presented here were taken with the relaxation method
\cite{Bachmann1972} based on an Oxford cryogenic system Maglab, in
which the heat capacity is determined by a direct measurement of
the thermal time constant, $\tau=(C_s+C_{add})/\kappa_w$, here
$C_s$ and $C_{add}$ are the heat capacity of the sample and
addenda, respectively, while $\kappa_w$ is the thermal conductance
between the chip and a thermal link. During the measurement the
sample was cooled to the lowest temperature under a magnetic field
( field-cooling ) followed by data acquisition in the warming up
process. The addenda includes a small sapphire chip, a tiny Cernox
temperature sensor, small amount of Wakefield thermal conducting
grease and gold leads ($\phi25\mu$m). All contributions to the
heat capacity from the addenda have been measured separately and
subtracted from the total specific heat before further analysis.

\begin{figure}[top]
\includegraphics[scale=1.1]{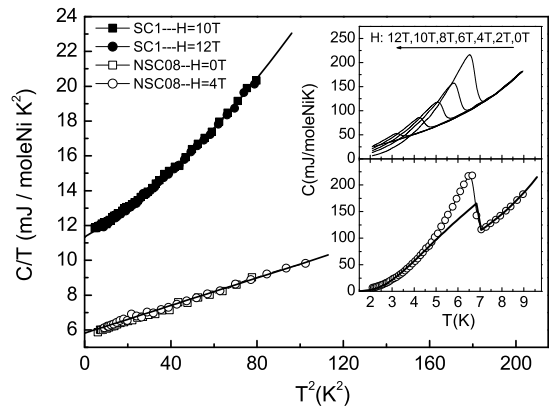}
\caption{\label{fig:fig2}A plot of $C/T$ vs $T^2$ for MgCNi$_{3}$ (SC1) and
MgC$_{0.85}$Ni$_{3}$ (NSC08) in various magnetic fields. The solid lines are fits to
Eq.~(\ref{eq:one}) as discussed in the text. The specific heat jump at the
superconducting transition temperature is completely depressed by the field above 10T.
Upper inset: The low-temperature specific heat for SC1 at various magnetic fields from 0T
to 12T. Lower inset: Fitting for the zero-field specific heat data (open circles).}
\end{figure}

The typical low temperature specific heat at various magnetic
fields up to 12 Tesla is presented in Fig.~\ref{fig:fig2}. For
sample NSC08, all the data fall into an universal curve,
indicating the independence of its specific heat on the magnetic
field. For sample SC1, all the data above $T_c$ are independent on
the field, and the normal state extends to the whole temperature
region investigated here when $H\geq 10T$. The upper critical
field $H_{c2}(T)$ can be determined from the position of the
specific-heat jump, and $H_{c2}(0)\approx 11T$ is then derived by
extrapolating $H_{c2}(T)$ to 0K. In order to describe the specific
heat data of NSC08 and the normal state data of SC1, we use the
following expression:
\begin{equation}
C_n(T)=\gamma_nT+\beta T^3+\delta T^5 \label{eq:one}
\end{equation}
where the linear-T term is the electronic contribution with $\gamma_n$ as Sommerfeld
parameter, the second term represents the phonon contribution according to the Debye
approximation, the last term is required to include deviations from the linear dispersion
of the acoustic modes in extended temperature range. The fitting results are presented in
Fig.~\ref{fig:fig2}, and the obtained parameters are shown in Table \ref{tab:table1}.
Besides the obvious difference between the Debye temperatures ($\Theta_D$) of SC1 and
NSC08 (due to the larger elastic modulus of NSC08 than that of SC1), a more prominent
difference is found between the Sommerfeld parameters $ \gamma_n$ of these two samples,
this is one of the main results in this work.

Before further analysis, we at first investigate the specific heat
data of the superconducting sample SC1 in zero field, as shown in
the lower inset of Fig.~\ref{fig:fig2}. The solid curve represents
the theoretical fits, the part above $T_c$ is the fitting result
to the normal state data of SC1 as mentioned above, while the part
below $T_c$ is an attempt to describe the data using the specific
heat expression of conventional superconductor:
$C_{es}=Aexp(-\Delta(0)/k_BT)/T^{1.5}$, in which $A$ is a function
of zero-temperature energy gap $\Delta(0)$ and the normal-state
DOS $g(0)$. This equation is a low temperature approximation and
is not adequate above $0.8T_c$ where $\Delta(T)$ has substantially
deviated from $\Delta(0)$. This conventional expression can fit
our data below $0.8T_c\approx 5K$ quite well, though the returned
$\Delta(0)$ (1.46meV) leads to a large value of
$2\Delta(0)/k_BT_c\approx 5$ comparing with the weak coupling
value 3.5. This indicates that MgCNi$_3$ is a strong coupling
superconductor, being consistent with the large value of $\Delta
C/\gamma_nT_c\approx 1.7$ ( the expected value for weak coupling
is 1.43 ), where $\Delta C$ is the specific heat jump at $T_c$.

\begin{table}
\caption{\label{tab:table1}Fits to Eq.~(\ref{eq:one}) for samples SC1 and NSC08.}
\begin{ruledtabular}
\begin{tabular}{ccccc}
sample & $\gamma_n$  & $\beta$  & $\delta$ & $\Theta_D$ \\
       & $mJ/molNiK^{2}$  & $mJ/molNiK^4$  & $mJ/molNiK^6$ & $K$ \\
\hline
SC1       & 11.3                &0.075        &0.00049    &351    \\
NSC08      & 5.8                 &0.040        &0          &434\\

\end{tabular}
\end{ruledtabular}
\end{table}

As shown above, the specific heat coefficient $\gamma_n$ of SC1 is
almost twice that of NSC08. In the frame of strong coupling
mechanism, given the value of $\gamma_n$ and $N(E_F)$, an
effective mass renormalization factor ($\lambda$) can be estimated
from the following relation \cite{MigdalA1958,RickayzenG1980}
\begin{equation}
\gamma_n=\frac{\pi^2}{3}k_B^2(1+\lambda)N(E_F) \label{eq:two}
\end{equation}
In order to obtain $N(E_F)$, we have calculated the electronic structures using a TB-LMTO
method based on Green function formalism \cite{TurekI1997} within the atomic sphere
approximation (ASA). The parameterized exchange-correlation potential of Ref
\cite{VoskoSH1980} and the scalar-relativistic Dirac equation \cite{KoellingDD1977} were
employed. The coherent potential approximation(CPA) used in our calculation allowed us to
study the electronic structure of MgC$_x$N$_{3}$ for any Carbon concentration. In the
crystal structure of cubic antiperovskite-type MgCN$_{3}$, the atoms occupy the positions
Mg (0,0,0), C (0.5,0.5,0.5), and Ni (0.5,0.5,0), (0.5,0,0.5), (0,0.5,0.5).
Self-consistency was reached using 1000 k-points within the irreducible wedge of the
simple cubic Brillouin zone. In the calculations, we adopted the experimental determined
\cite{AmosTG2002} relation between the lattice parameter $a$ and the carbon concentration
$x$. A fixed radii proportion as $2.078:1:1.617$ (Mg:C:Ni) was taken for the
stoichiometric situation ($x=1$), and only the radii of carbon-site was reduced with
decreasing $x$.

\begin{figure}
\includegraphics[scale=1.1]{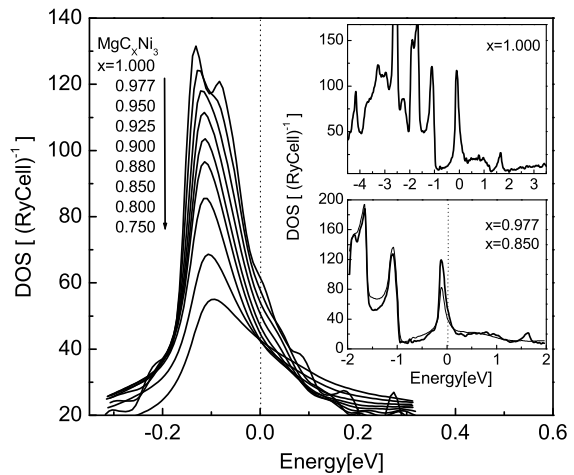}
\caption{\label{fig:fig3}Electronic DOS of MgC$_{x}$Ni$_3$ for various carbon
concentration calculated within the ASA. The dotted line indicate the Fermi level. Upper
inset: Electronic DOS of MgCNi$_3$ in a lager energy scale. Lower inset: Redistribution
of DOS with the change of carbon concentration. The thick solid line and thin one
corresponds to $x=0.977$ and $0.85$, respectively.}
\end{figure}

The DOS of MgC$_x$N$_{3}$ for $x=1$ are presented in the upper
inset of Fig.~\ref{fig:fig3}, which is consistent with the reports
of other groups \cite{SinghDJ2001,ShimJH2001,SheinIR2002,Kim2002}.
The $\pi^{*}$ antibonding states of Ni-$3d$ and C-$2p$ are located
just below $E_F$, yielding a high DOS peak whose height is proved
to be sensitive to the exact method. Fortunately, the obtained
value of $N(E_F)=4.53$ eV$^{-1}$cell$^{-1}$ is not far away from
the result of general potential linearized augmented planewave
method (4.99 eV$^{-1}$cell$^{-1}$) \cite{SinghDJ2001} and that of
full-potential LMTO method (4.57 eV$^{-1}$cell$^{-1}$)
\cite{SheinIR2001}, which is a good starting point for the
following consideration of the doping effect.

\begin{figure}
\includegraphics[scale=1.1]{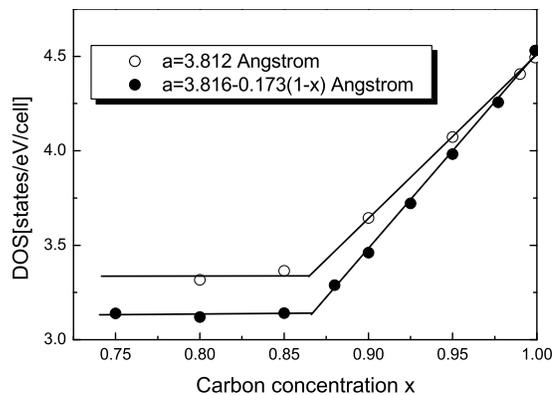}
\caption{\label{fig:fig4}The dependence of $N(E_F)$ of MgC$_{x}$Ni$_3$ on carbon
concentration. $N(E_F)$ for fixed $a=3.812\AA$ is also calculated as a comparison. The
solid lines are guides to eyes.}

\end{figure}

Fig.~\ref{fig:fig3} shows the evolution of the DOS around Fermi level with the change of
carbon concentration $x$ in MgC$_{x}$Ni$_3$. It is found that the Fermi level slightly
shifts towards low energy with the decreasing $x$, while the Ni$3d$-C$2p$ antibonding
peak is depressed remarkably. In the nearest-neighbor approximation, the degeneration of
this DOS peak indicate that the carbon-site vacancies locally break the hybridization
between Ni-$3d$ and C-$2p$ orbitals and hence lead to a redistribution of the electronic
states. Atom-resolved DOS shows that the near-$E_F$ DOS peak of Ni-$3d$ component and
that of C-$2p$ component do not depart from each other with decreasing $x$, indicating
that carbon-site vacancies does not destroy the Ni$3d$-C$2p$ bond for all carbon
concentrations investigated here. Therefore, the decrease of the peak possibly means the
enhanced itinerancy of electrons. As shown in Fig.~\ref{fig:fig3}, the depressed DOS peak
just below $E_F$ results in a notable reduction of $N(E_F)$. The dependence of $N(E_F)$
on the carbon concentration is presented in Fig.~\ref{fig:fig4}. We can see a obvious
decrease of $N(E_F)$ from 4.26eV$^{-1}$ at $x=0.977$ to 3.14eV$^{-1}$ at $x=0.85$.
Moreover, the doping dependence of $N(E_F)$ is linear above a critical doping
$x_c\approx0.88$ which is very close to the carbon content in NSC08 and also the lower
limit of the carbon concentration for SC in MgC$_{x}$Ni$_3$ \cite{AmosTG2002}. Below
$x_c$, the doping dependence of $N(E_F)$ becomes much weaker than that above $x_c$.

By inserting the experimental value of $\gamma_n$ and the
calculated $N(E_F)$ into Eq(~\ref{eq:two}), we can determine the
coupling constant as $\lambda=2.37$ and 1.35 for the samples SC1
and NSC08, respectively. The former is close to that reported by
W\"alte {\it et al.} \cite{WalteA2002}, while the remarkable
difference in the coupling strength between a superconducting
sample and a carbon-deficient non-superconducting one is for the
first time found in the MgC$_{x}$Ni$_3$ system. However, applying
$\lambda=2.37$ to the McMillan's $T_c$ formula
\cite{McMillanWL1968}
$T_c=\frac{\Theta_D}{1.45}e^{-\frac{1+\lambda}{\lambda}}$ gives
rise to $T_c>30K$, which seems to be too large as compared to
experimental value $T_c=7K$. In other words, $T_c=7K$ corresponds
to $\lambda=0.42$, which is much smaller than the value determined
here. This contradiction can be resolved if we adopt the picture
that there are two kinds of boson-mediated electron-electron
interactions, for example, electron-phonon coupling (EPC) and spin
fluctuation (SF) \cite{SinghDJ2001,WalteA2002}. Due to the
different effect of the spin fluctuations on mass renormalization
and superconducting properties, it should add in the mass
renormalization term and subtract in the pairing term for SC. So
in the frame of strong coupling, the specific heat determined
$\lambda$ has two contributions:
$\lambda=\lambda_{ph}+\lambda_{spin}$. Similarly, the simplified
McMillan's $T_c$ formula should be modified as \cite{SinghDJ2001}
\begin{equation}
T_c=\frac{\Theta_D}{1.45}e^{-\frac{1+\lambda_{ph}+\lambda_{spin}}{\lambda_{ph}-\lambda_{spin}}}
\label{eq:three}
\end{equation}
Using the determined parameters for sample SC1 (namely, $\lambda=2.2$, $T_c=7K$ and
$\Theta=351K$), one obtains $\lambda_{ph}-\lambda_{spin}=0.95$ from Eq.(~\ref{eq:three}).
Combining this result with $\lambda_{ph}+\lambda_{spin}=2.37$, we determine the strength
of both EPC and electron-paramagnon coupling as $\lambda_{ph}=1.66$ and
$\lambda_{spin}=0.71$. Firstly, by assuming that $\lambda_{spin}$ is invariable with the
decreasing carbon concentration, we could obtain $\lambda_{ph}=0.64$ for sample NSC08.
This suggests a strong dependence of the EPC strength on the carbon concentration in
MgC$_x$Ni$_3$. For NSC08, $\lambda_{ph}=0.64$ has become smaller than
$\lambda_{spin}=0.71$, which is closely associated with the disappearance of SC.
Secondly, on the assumption that $x=0.85$ is the critical stoichiometry for SC, i.e.,
$\lambda_{ph}=\lambda_{spin}$, we obtain $\lambda_{ph}=\lambda_{spin}=0.665$. This
indicates a $6\%$ decrease of $\lambda_{spin}$ from $x=0.977$ to $x=0.85$. However,
$\lambda_{ph}$ is lost near $60\%$ in the same system, which seems to be responsible for
the suppression of SC in MgC$_x$Ni$_3$.

We have been aware of some recent calculations \cite{SheinIR2002,SheinIR2001} suggesting
that both Co doping and Cu doping in MgCNi$_3$ lead to a reduction of $N(E_F)$, which
seems to be critical in depressing SC in this system (MgCNi$_{3-x}$Y$_x$ [Y=Co or Cu]).
However, these investigations were based on the ordered substitutions, i.e., x=1, 2 or 3,
which is somewhat away from the key problem because the solubility of Cu is limited to
$3\%$ ($x=0.1$) in technology and the bulk SC disappears for only $1\%$ Co doping
($x=0.03$). Our further calculations did not show obvious reduction of $N(E_F)$ in this
doping range. Therefore, some necessary investigations on the variation of the EPC in
this system must be supplemented.

In summary, we have investigated the specific heat data in
MgC$_x$Ni$_3$ system and found a remarkable difference between the
Sommerfeld parameters of a superconducting sample ($x\approx1$)
and a non-superconducting one ($x\approx 0.85$). Band calculations
reveal a distinct decrease of the density of the electronic states
at Fermi level with the decreasing carbon concentration. Further
analysis indicate that there is more than one kind of
boson-mediated electron-electron interactions existing in
MgC$_x$Ni$_3$ system. In the frame of strong coupling theory, a
substantial depression of the electron-phonon coupling caused by
the decrease of carbon concentration is for the first time found
in this system and seems to be responsible for the impairment of
its superconductivity.

% If you have acknowledgments, this puts in the proper section head.
\begin{acknowledgments}
% put your acknowledgments here.
This work is supported by the National Science Foundation of China (NSFC 19825111,
10074078), the Ministry of Science and Technology of China ( project:
NKBRSF-G1999064602 ), and Chinese Academy of Sciences with the Knowledge Innovation
Project.
\end{acknowledgments}

% Create the reference section using BibTeX:
%\bibliography{NoEndingPoint}

\end{document}